\documentclass{PoS}

\title{Improved analysis of the CLFV decay of muonic atoms $\mu^-e^-\rightarrow e^-e^-$}

\ShortTitle{Improved analysis of the CLFV decay of muonic atoms $\mu^-e^-\rightarrow e^-e^-$}

\author{\speaker{Yuichi Uesaka},$^a$ Yoshitaka Kuno,$^a$ Joe Sato,$^b$ Toru Sato$^a$ and Masato Yamanaka$^c$ \\
        \llap{$^a$}Department of Physics, Osaka University, Toyonaka, Osaka 560-0043, Japan\\
        \llap{$^b$}Physics Department, Saitama University, 255 Shimo-Okubo, Sakura-ku, Saitama 338-8570, Japan\\
        \llap{$^c$}Department of Physics, Nagoya University, Nagoya 464-8602, Japan\\
        E-mail: \email{uesaka@kern.phys.sci.osaka-u.ac.jp}, \email{kuno@phys.sci.osaka-u.ac.jp}, \email{joe@phy.saitama-u.ac.jp}, \email{tsato@phys.sci.osaka-u.ac.jp}, \email{yamanaka@eken.phys.nagoya-u.ac.jp}}

\abstract{Koike \textit{et al.} proposed the charged lepton flavor violation (CLFV) decay of the muonic atom $\mu^-e^-\rightarrow e^-e^-$ as one of the promising processes to search for new physics beyond the standard model \cite{Koike2010}.
It was found that the attractive interaction of leptons with the nucleus enhances the transition rate of the $\mu^-e^-\rightarrow e^-e^-$ process.
We report on our improved analysis of this process by taking into account the distortion of the out-going electrons in the nuclear Coulomb potential and the relativistic treatment of the muon and the electrons.
As results, we found significant enhancement of the transition rate.
The transition rate for $^{208}$Pb becomes about 7 times larger than the previous estimation, which enhances the sensitivity of this process to discover the CLFV interaction.
We also report on the energy spectrum of the out-going electron.}

\FullConference{Flavor Physics \& CP Violation 2015,\\
		May 25-29, 2015\\
		Nagoya, Japan}

\usepackage{amsmath,bm}
\usepackage{braket}

\begin{document}

\section{Introduction}
A new process, $\mu^-e^-\rightarrow e^-e^-$ transition in a muonic atom has been proposed by Koike et al. \cite{Koike2010} in 2010 for the search of the charged lepton flavor violation (CLFV) process.
This process has special features that (1) One can probe both contact and photonic interactions similar to $\mu^+\rightarrow e^+e^+e^-$.
(2) The experimental signal is expected to be clear, since the
total energy of final electrons is uniquely given by the sum of muon and electron masses minus binding energies of initial bound leptons.
(3) A larger transition probability for heavier muonic atom is expected due to enhancement of the overlap of lepton wave functions due to attractive nuclear Coulomb interaction.

In Ref. \cite{Koike2010}, the transition rate was evaluated by using the non-relativistic bound state wave functions of muon and electron and the plane wave approximation of the final electrons and it was shown the transition rate increases with the atomic number $Z$ as $\Gamma \sim Z^3$.
In order to disentangle the mechanism of CLFV interaction, qualitative evaluation of the transition process is necessary and therefore it is important to update the finding of Ref. \cite{Koike2010}.
The importance of the Coulomb distortion for the $\mu^--e^-$ conversion process of muonic atom has been reported in \cite{Kitano2002,Shanker1979,Czarnecki1998}.
In this report we investigate the effects of Coulomb interactions of the relativistic leptons.
We have developed a formalism of the CLFV transition with the partial wave expansion of leptons as used for the nuclear beta decay and muon capture reaction \cite{Koshigiri1979}.
In section \ref{sec:Formalism}, we briefly describe the relevant effective Lagrangian for the $\mu^-e^-\rightarrow e^-e^-$ process and develop a formula of the transition rate using the partial wave expansion of the lepton wave function.
Our estimations of the transition rate for $\mu^-e^-\rightarrow e^-e^-$ in the muonic atom is presented in section \ref{sec:Results}.
Finally we summarize the report in section \ref{sec:Summary}.

\section{Formalism \label{sec:Formalism}}
The effective Lagrangian which describes the CLFV transition $\mu^-e^-\rightarrow e^-e^-$ in muonic atoms has two parts:
\begin{eqnarray}
\mathcal{L}_I&=&\mathcal{L}_\mathrm{contact}+\mathcal{L}_\mathrm{photonic},
\end{eqnarray}
which generate a short-range four fermion interaction $\mathcal{L}_\mathrm{contact}$ and a long-range photon-exchange interaction $\mathcal{L}_\mathrm{photonic}$, respectively.
In this report, we concentrate on the contact interactions as our first attempt to examine the role of Coulomb interaction on the CLFV transition in muonic atom.
Following \cite{Koike2010,Kuno2001}, the contact interaction is given as
\begin{eqnarray}
\mathcal{L}_\mathrm{contact}&=&-\frac{4G_F}{\sqrt{2}}[g_1(\overline{e_L}\mu_R)(\overline{e_L}e_R)+
g_2(\overline{e_R}\mu_L)(\overline{e_R}e_L) \nonumber\\
&&+g_3(\overline{e_R}\gamma_\mu\mu_R)(\overline{e_R}\gamma^\mu\mu_R)+g_4(\overline{e_L}\gamma_\mu\mu_L)
(\overline{e_L}\gamma^\mu\mu_L) \nonumber\\
&&+g_5(\overline{e_R}\gamma_\mu\mu_R)(\overline{e_L}\gamma^\mu\mu_L)+g_4(\overline{e_L}\gamma_\mu\mu_L)
(\overline{e_R}\gamma^\mu\mu_R)]+[h.c.],
\end{eqnarray}
where $G_F$ is the Fermi coupling constant, and $g_i$'s ($i=1,2,\cdots,6$) are dimensionless coupling constants.

The transition rate $\Gamma_0$ was studied from the following formula in Ref. \cite{Koike2010},
\begin{eqnarray}
\Gamma_0=\sigma_{\mu e\rightarrow ee}v_\mathrm{rel}\left|\psi_e(0)\right|^2\propto(Z-1)^3,
\end{eqnarray}
where $\sigma_{\mu e\rightarrow ee}$ is the cross section of $\mu^-e^-\rightarrow e^-e^-$, $v_\mathrm{rel}$ is a relative velocity between the muon and electron, and $\psi_e(0)$ is the bound electron wave function at the origin.
The formula was derived by using the non-relativistic wave function of bound lepton and plane wave approximation for the final electrons.
They found the transition rate increases as $Z^3$ for heavy atom.

In this report we examine the role of Coulomb interaction of leptons on the transition rate.
We start from the standard formula of transition rate assuming the two electron is emitted by CLFV interaction between the bound muon and electron as
\begin{eqnarray}
\Gamma=2\pi\sum_{f}\sum_{\overline{i}}\delta(E_f-E_i)\left|\braket{\psi^e_{\vec{p}_1}\psi^e_{\vec{p}_2}|L_I|\psi^\mu_B\psi^e_B}\right|^2.
\end{eqnarray}
We use the Coulomb scattering wave function of the Dirac equation for the final two electrons and the relativistic bound state wave function for the muon and electron.
The Coulomb potential from finite size charge distribution of nuclei is used.
After the partial wave expansion of the scattering wave function and straightforward calculations, we arrive at a formula
\begin{align}
\Gamma=&\frac{G_F^2}{\pi^3}\int_{m_e}^{m_\mu-B_\mu-B_e}dE_{p_1}|\bm{p}_1||\bm{p}_2|\sum_{J,\kappa_1,\kappa_2}(2J+1)(2j_{\kappa_1}+1)(2j_{\kappa_2}+1)\left|\sum_{i=1}^{6}g_iW_i(J,\kappa_1,\kappa_2,E_{p_1})\right|^2.
\label{eq:result}
\end{align}
Here $W_i$ includes the overlap integral of lepton wave functions.
A explanation and derivation of the formula Eq. (\ref{eq:result}) is given in Ref.\cite{Uesaka2015}.

\section{Results \label{sec:Results}}
\begin{figure}[htb]
 \begin{minipage}{0.47\hsize}
	\centering
	\includegraphics[width=.8\textwidth]{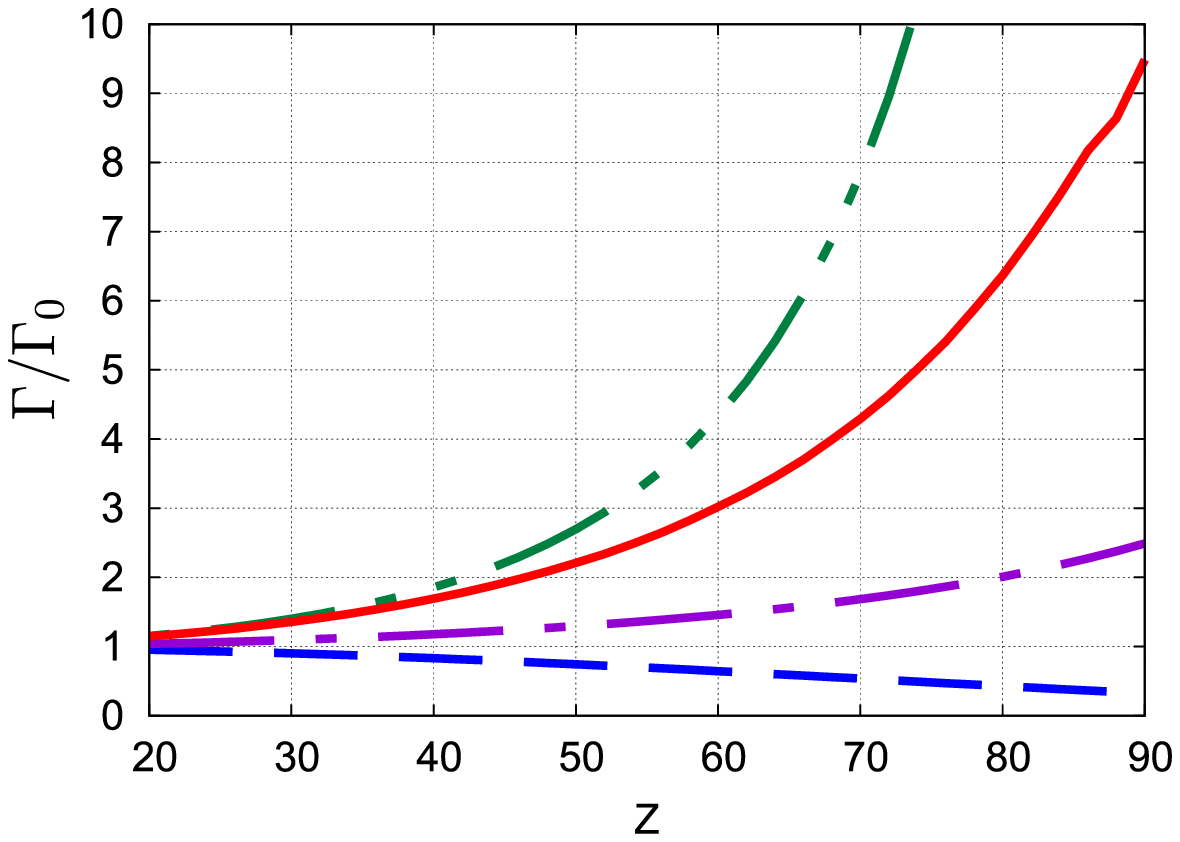}
	\caption{The atomic number ($Z$) dependence of the ratio of the decay rate $\Gamma/\Gamma_0$.}
	\label{fig:decayrate}
 \end{minipage}%
 \begin{minipage}{0.06\hsize}
 \hspace{0.06\hsize}
 \end{minipage}%
 \begin{minipage}{0.47\hsize}
	\centering
	\includegraphics[width=.8\textwidth]{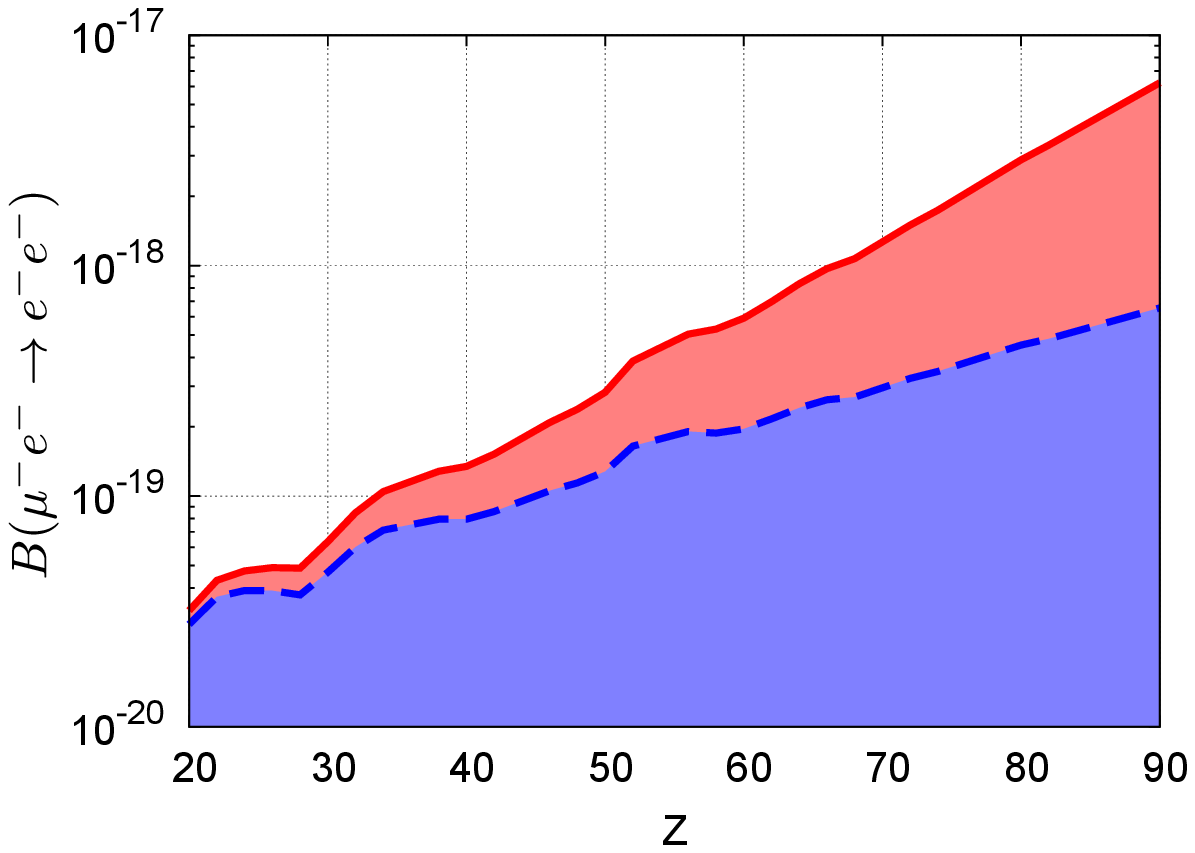}
	\caption{Upper limits on $Br(\mu^-e^-\rightarrow e^-e^-)$,
	imposed by the experimental upper limits of $Br(\mu^+\rightarrow e^+e^+e^-)$.}
	\label{fig:branchingratio}
 \end{minipage}
\end{figure}

In the following results, we use only the $g_1$ term and set $g_2=g_3=g_4=g_5=g_6=0$ for simplicity.
The role of Coulomb effects for the other terms is essentially the same as $g_1$ term.
In Figure \ref{fig:decayrate}, the ratio $\Gamma/\Gamma_0$ is shown, where deviation of the ratio from $1$ shows the role of Coulomb distorted wave.
At first, the transition rate is evaluated by using the same lepton wave functions as the previous work, i.e. the non-relativistic bound states and plane wave scattering states in the dashed curve in Figure \ref{fig:decayrate}.
The ratio deviates from $1$ for larger $Z$, since the finite size of both muon and electron bound states taken into account in our formula is more important for high $Z$ atom.
When we replace the final plane wave electrons with the Dirac wave function of point nuclear charge, the transition rate increases as shown in the dot-dashed curve.
The enhancement of the transition rate is more pronounced when we use the relativistic wave functions with the point nuclear charge for both bound states and scattering states shown in the dash-two-dotted curve as we can expect from the transition density.
A more realistic description of $\Gamma/\Gamma_0$ is obtained by using the uniform nuclear charge distribution shown in the solid curve in Figure \ref{fig:decayrate}.
The results show that  the $Z$ dependence of the $\Gamma$ is stronger than $(Z-1)^3$.
The approximate $\Gamma_0$ is indeed a good approximation  for smaller $Z\sim 20$.
The ratio $\Gamma/\Gamma_0$ is about $7.0$ for the $^{208}$Pb.

Then we evaluate the upper limits of the branching ratio of $\mu^-e^-\rightarrow e^-e^-$ in a muonic atom, defined as
\begin{align}
Br(\mu^-e^-\rightarrow e^-e^-)\equiv\tilde{\tau}_\mu\Gamma(\mu^-e^-\rightarrow e^-e^-),
\end{align}
where $\tilde{\tau}_\mu$ is a lifetime of a muonic atom.
Since only the available information is the upper limit of the $\mu^+\rightarrow e^+e^+e^-$ decay branching ratio, we estimate the maximum strength of the CLFV interaction assuming the short-range CLFV interaction from the upper limit of the branching ratio.
The dashed (blue) curve in Figure \ref{fig:branchingratio} shows the result of previous work \cite{Koike2010}.
The increased branching ratio for large $Z$ atom is obtained from our results shown in the solid (red) curve.
The branching ratio reaches about $10^{-17}$ for a large $Z$ atom.

\section{Summary \label{sec:Summary}}
We have estimated the transition rate for $\mu^-e^-\rightarrow e^-e^-$ in muonic atoms due to the CLFV interaction and we have shown the importance of the Coulomb distortion for the emitted electrons and the relativistic wave function of bound lepton.
The transition rates are largely enhanced for large $Z$ atom compared with the previous estimation.

In the next work, we plan to analyze the long-range photon exchange CLFV interaction.
Besides, it would be interesting to investigate a possibility to discriminate various forms of CLFV interactions by using the observables of $\mu^-e^-\rightarrow e^-e^-$.
Examples are an energy spectrum of a emitted electron, the angular correlation between two emitted electrons, and helicities of electrons.


\begin{thebibliography}{99}
\bibitem{Koike2010}
M. Koike, Y. Kuno, J. Sato and M. Yamanaka,
Phys. Rev. {\bf 105}, 121601 (2010).

\bibitem{Kitano2002}
R. Kitano, M. Koike, and Y. Okada,
Phys. Rev. D {\bf 66}, 096002 (2002).

\bibitem{Shanker1979}
O. Shanker,
Phys. Rev. D {\bf 20}, 1608 (1979).

\bibitem{Czarnecki1998}
A. Czarnecki, W. J. Marciano and K. Melnikov, 
arXiv:hep-ph/9801218 (1998).

\bibitem{Koshigiri1979}
K. Koshigiri, N. Nishimura, H. Ohtsubo and M. Morita,
Nucl. Phys. A{\bf 319} 301, (1979).

\bibitem{Kuno2001}
Y. Kuno and Y. Okada,
Rev. Mod. Phys. {\bf 73}, 151 (2001).

\bibitem{Uesaka2015}
Y. Uesaka, Y. Kuno, J. Sato, T. Sato and M. Yamanaka, in preparation.

\end{thebibliography}
\end{document}